\documentstyle[aps,prb,twocolumn,floats,psfig]{revtex}

\begin{document}

\bibliographystyle{prsty}

\title{ 
\begin{flushleft}
{\small \em submitted to}\\
{\small 
PHYSICAL REVIEW B 
\hfill
VOLUME XX, %{\normalsize XX}, 
NUMBER X % {\normalsize X} 
$\qquad$
\hfill 
MONTH XXX
%DECEMBER,  {\normalsize 6499$-$6502} 
}
\end{flushleft}  
Effective classical Hamiltonians for spin systems: \\
A closed form with quantum spin-wave effects
\vspace{-1mm}
} 
\author{
D. A. Garanin \cite{e-gar}, D. V. Dmitriev \cite{e-dmi}, and P. Fulde 
}
\address{
Max-Planck-Institut f\"ur Physik komplexer Systeme, N\"othnitzer Strasse 38,
D-01187 Dresden, Germany 
\\
%}
%\date{\today}
%\maketitle
%\abstract{
\smallskip
{\rm(Received 15 April 2001)}
\bigskip\\
\parbox{14.2cm}
{\rm
Thermodynamic properties of any quantum spin system can be described by the formally exact, although in general intractable, effective classical Hamilton function ${\cal H}$.
Here we obtain an explicit form of ${\cal H}$ which applies at $T\ll J S^2$, where $J$ is the exchange and $S$ in the spin value, and incorporates quantum effects at the level of the spin-wave theory (SWT).
For a quantum Hamiltonian $\hat H$ of Heisenberg form, ${\cal H}$ is also Heisenberg but with a long-range effective exchange $J_{ij}^{\rm eff}$, which is the price for including quantum effects.
For three-dimensional magnets, classical SWT with ${\cal H}$ yields the same results as quantum SWT for the original system, in the antiferromagnetic case with the $1/S$ correction to the ground-state energy.
For nontrivial one- and two-dimensional systems, reduction of the problem from quantum to effective classical allows to apply such methods as classical Monte Carlo simulation and the $1/D$ expansion, where $D$ is the number of spin components, etc. 
\smallskip
\begin{flushleft}
\end{flushleft}
} 
} 
\maketitle

\section{Introduction}
\label{sec_introduction}

The partition function of any quantum spin system with a Hamiltonian $\hat H$ can be reduced to that of an {\em effective classical} spin system with the Hamilton function ${\cal H}$ by rewriting it in the basis of spin coherent states $|{\bf n}_i\rangle$, the unit vectors ${\bf n}_i$ serving as classical spin vectors in ${\cal H}$. \cite{lie73,klafulgar99,garklaful00}
The potential significance of the possibility to treat quantum systems as
classical is quite apparent: In one and two dimensions one can perform
classical Monte Carlo (MC) simulations, and use the classical-spin diagram technique\cite{garlut84d,gar96prb} or the $1/D$ expansion, where $D$ is the number of spin components,\cite{gar94jsp,gar96jsp}  etc.  
However, since ${\cal H}$ incorporates all quantum effects of the system, in most cases its formally exact expression cannot be brought into a practically applicable form without approximations.

In Ref.\ \onlinecite{klafulgar99} a {\em cumulant expansion} (CE) was applied to semiclassical ($S\gg 1$) one- and two-spin systems to obtain a tractable form of the effective classical Hamilton function ${\cal H}$.
In Ref.\ \onlinecite{garklaful00} this approach has been generalized to many-body spin systems.
The cumulant expansion yields ${\cal H}$ as a series of non-Heisenberg multispin terms of increasing complexity, the zero-order term being the classical counterpart ${\cal H}^{(0)}$ of the quantum Hamiltonian $\hat H$.
CE is an expansion in powers of $JS/T$ with additional terms of higher orders in $1/S$.
Thus it diverges for $T \lesssim JS$, where $JS$ is of the order of the maximal spin-wave energy $\varepsilon_{\bf k}^{\rm max}$ and is much smaller than another energy scale $JS^2$ which is of the order of the Curie temperature $T_c$ for 3$d$ systems.
Note that in the applicability region of the cumulant expansion the boson occupation numbers $n_{\bf k} \equiv 1/(e^{\beta\varepsilon_{\bf k}} -1)$ can be expanded in $\beta\varepsilon_{\bf k} \ll 1$, that is, the system behaves classically with quantum corrections.
Note that one should not confuse the cumulant expansion with the high-temperature series expansion (HTSE) which is an expansion in $JS^2/T$ and for $S\gg 1$ has a much narrower applicability range than the former.
 
The presence of the two different energy scales in the problem suggests continuing the results of Ref.\ \onlinecite{garklaful00} into the region $T\lesssim JS$, which is the aim of this work.
Indeed, at the convergence limit of the cumulant expansion, $T\sim JS$, there is already a strong short-range order (which is formed at $T\ll JS^2$), i.e., fluctuations of spin vectors coupled by terms of the CE, relative to each other, are small.
Thus the form of CE should simplify for $T\ll JS^2$, and one expects that the resulting series of effective Hamilton functions can be summed up to yield some ${\cal H}$ that is valid down to $T=0$.
As we shall see, this works indeed for the ferromagnetic model, where the information contained in ${\cal H}^{(1)}$ and ${\cal H}^{(2)}$ of Ref.\ \onlinecite{garklaful00} is sufficient to guess all other terms of CE and to sum them up.

For more complicated models, such as antiferromagnets, higher-order terms of the simplified CE cannot be guessed from the first few ones, therefore one has to search for an alternative approach.
The latter can be found. It turns out to be more satisfactory since one does not have to expand ${\cal H}$ in a power series and then sum it up again.
Instead it makes use of the Holstein-Primakoff expansion of spin operators up to bilinear boson terms, which is familiar from the SWT.  
Thus our result for ${\cal H}$ will be a good approximation in all cases where SWT works, i.e., for most of magnetic models except for those where quantum effects lead to nontrivial ground states.

The paper is organized as follows.
In Sec.\ \ref{sec_cumtosw}  derivation of the general formula for the effective classical Hamilton function ${\cal H}$ is given. 
Thereby the spin coherent states are used and the results of the cumulant expansion of Ref.\ \onlinecite{garklaful00} are recapitulated.
The CE for ferromagnets is simplified for $T\ll JS^2$ and summed up to yield ${\cal H}$ which is valid down to $T=0$.
In Sec.\ \ref{sec_sw} the Holstein-Primakoff expansion is used to obtain the expressions for ${\cal H}$ for ferro- and antiferromagnetic models.
In Sec.\ \ref{sec_discussion} possible applications of the results obtained are discussed.

\section{Effective classical Hamiltonians: From cumulant expansion to spin waves}
\label{sec_cumtosw}

\subsection{Main idea and basic relations}

Thermodynamic properties of any quantum spin system with a Hamiltonian $\hat H$ can be described in an alternative way via the {\em effective classical} Hamilton function ${\cal H}$ which can be constructed using spin  coherent states $|{\bf n}_i\rangle$ on lattice sites $i$. \cite{lie73,klafulgar99,garklaful00}
The vectors ${\bf n}_i$ parametrizing these maximum-projection states serve as classical spin vectors in ${\cal H}$, and the partition function ${\cal Z}$ of the quantum spin system can be calculated as an integral over $d{\bf n}_i$.
The derivation of this effective classical formalism for spin systems is based on the representation of the unit operator in the (overcomplete) basis of spin coherent states
%\marginpar{ResUnity}
%
\begin{equation}\label{ResUnity}
{\bf 1} = \frac{2S+1}{4 \pi} \int d{\bf n}\, |{\bf n}\rangle \langle{\bf n}|. 
\end{equation}
For a single-spin system, the trace of an operator $\hat A$ over any complete orthonormal basis $|m\rangle$ becomes \cite{lie73}
%\marginpar{TraceRewrite}
%
\begin{eqnarray}\label{TraceRewrite}
&&
{\rm tr}\, \hat A = \sum_m \langle m| \hat A | m \rangle 
= \frac{2S+1}{4 \pi} \int d {\bf n} 
\sum_m \langle m|\hat A|{\bf n}\rangle \langle {\bf n}|m\rangle 
\nonumber\\
&&\qquad\qquad
= \frac{2S+1}{4 \pi} \int d {\bf n}\, \langle {\bf n}| \hat A| {\bf n}\rangle.
\end{eqnarray}
Similarly, the partition function for a many-spin quantum Hamiltonian $\hat H$ can be written as 
%\marginpar{ZCohState}
%
\begin{equation}\label{ZCohState}
{\cal Z} =   
\left(\frac{2S+1}{4 \pi} \right)^N \int \prod_{i=1}^N d {\bf n}_i\, 
\langle \{ {\bf n}_i \}| e^{- \beta \hat H} |\{ {\bf n}_i \}\rangle,   
\end{equation}
where $\beta \equiv 1/T$.
It has the same form as the one for {\em classical} systems, provided one defines the effective classical Hamilton function ${\cal H}$ by the formula
%\marginpar{HClass}
%
\begin{equation}\label{HClass}
e^{-\beta {\cal H}(\{ {\bf n}_i \})} = \langle  \{ {\bf n}_i \}| e^{- \beta \hat H} | \{ {\bf n}_i \} \rangle.
\end{equation}

The function ${\cal H}$ evidently depends on temperature, thus calculation of physical quantities should be done with care.
For instance, the internal energy is given by 
\begin{equation}\label{Udef}
U = - \frac { \partial \ln {\cal Z} } { \partial \beta } = 
\left \langle \frac { \partial (\beta {\cal H} ) }{ \partial \beta } \right\rangle , 
\end{equation}
where $\langle \ldots \rangle$ denotes a classical thermal average with ${\cal H}$.
One can see that ${\cal H}^* \equiv \partial (\beta {\cal H} )/ \partial \beta \neq {\cal H}$.
The same care should be taken in calculation of the magnetization and correlation functions (see Ref.\ \onlinecite{garklaful00}).

The matrix element over the spin coherent states which defines ${\cal H}$ in Eq.\ (\ref{HClass}) can be calculated explicitly only for the simplest models such as one spin in a field. \cite{klafulgar99}
For many-body systems such as the Heisenberg model
%\marginpar{QHam}
%
\begin{equation}\label{QHam}
\hat H = - \sum_i {\bf H}_i \cdot {\bf S}_i - \frac 12 \sum_{ij} J_{ij} {\bf S}_i \cdot 
{\bf S}_j ,
\end{equation}
one has to resort to approximate methods. 
A plausible idea is to solve the problem perturbatively in $1/S$ in the quasiclassical case $S\gg 1$.
In the classical limit, $S\to\infty$, one obtains the classical Hamilton function
%\marginpar{ClassHam}
%
\begin{equation}\label{ClassHam}
{\cal H}^{(0)} = -  \sum_i {\bf h}_i \cdot {\bf n}_i - \frac 12 \sum_{ij} \tilde J_{ij} {\bf n}_i 
\cdot {\bf n}_j, 
\end{equation}
as was rigorously proved by Lieb. \cite{lie73}
Here
%\marginpar{RedVar}
%
\begin{equation}\label{RedVar}
{\bf h}_i \equiv S {\bf H}_i,   \qquad  \tilde J_{ij} \equiv S^2 J_{ij}  
\end{equation}
are the reduced magnetic field and the exchange interaction, respectively.

Quantum corrections to Eq.\ (\ref{ClassHam}) were calculated in Ref.\ \onlinecite{garklaful00} with the help of a cumulant expansion which yields a series of multi-site non-Heisenberg Hamilton functions of increasing complexity.
These expressions are too lengthy for reproducing them here.
The small parameter required for the validity of the cumulant expansion, in addition to the obvious
$1/S \ll 1$, is  
%\marginpar{CumExpPar}
%
\begin{equation}\label{CumExpPar}
 \frac { JS}{ T} \ll 1,  
\end{equation}
where $J$ is the coupling constant when only nearest neighbor interactions are
taken into account.
The inequality breaks down at a temperature $T \sim  JS$ which is of the order of the maximal energy of spin waves.
On the other hand, a strong short-range order sets for $T \lesssim  JS^2$, which is much higher than $JS$ for $S\gg 1$.
Thus for $S\gg 1$ the results of the cumulant expansion should simplify in the temperature range $JS \ll T \ll JS^2$, and one expects that CE can be summed up in this range to yield an expression for ${\cal H}$ which is valid down to $T=0$. 
This is the idea of the present work which in the following subsection is presented for ferromagnets.

\subsection{Effective low-temperature Hamilton function for ferromagnets}

The non-Heisenberg terms of the cumulant expansion for the Hamilton function ${\cal H}$ (see Ref.\ \onlinecite{garklaful00}) can be simplified for $T \ll \tilde J = JS^2$ since there is strong short-range order in the system in this temperature range.
For the ferromagnetic model which we will deal with in this subsection, the scalar product ${\bf n}_i \cdot {\bf n}_j$ is close to unity. 
Thus one can neglect the terms quadratic in $1-{\bf n}_i \cdot {\bf n}_j$ and simplify
%\marginpar{Simplify}
%
\begin{eqnarray}\label{Simplify}
&&
({\bf n}_i \cdot {\bf n}_j) ({\bf n}_j \cdot {\bf n}_l) =
[ 1 + ({\bf n}_i \cdot {\bf n}_j) -1 ] [ 1 + ({\bf n}_j \cdot {\bf n}_l) - 1 ]
\nonumber\\
&&\qquad
\cong -1 + ({\bf n}_i \cdot {\bf n}_j) + ({\bf n}_j \cdot {\bf n}_l)
\end{eqnarray}
With the help of this formula, the first-order Hamilton function in zero field,
%\marginpar{H1}
% 
\begin{equation}\label{H1}
{\cal H}^{(1)} = 
  -\frac \beta {4S} \sum_{ijl} \tilde J_{ij} \tilde J_{jl} 
[ ({\bf n}_i \cdot {\bf n}_l) - ({\bf n}_i \cdot {\bf n}_j) ({\bf n}_j \cdot {\bf n}_l) ]
\end{equation}  
which was derived before [see Eq.\ (20) of Ref.\ \onlinecite{garklaful00}], 
simplifies to
%\marginpar{H1simp}
% 
\begin{equation}\label{H1simp}
{\cal H}^{(1)} \cong 
  -\frac {\beta \tilde J_0} {4S} \sum_{ij} V_{ij}^{(1)}{\bf n}_i \cdot {\bf n}_j 
\end{equation}  
with
%\marginpar{V1def}
% 
\begin{equation}\label{V1def}
V_{ij}^{(1)} \equiv \delta_{ij} \tilde J_0 - 2\tilde J_{ij} + \frac 1 {\tilde J_0 } 
\sum_l \tilde J_{il} \tilde J_{lj}.
\end{equation}  
The second-order terms of the Hamilton function of Ref.\ \onlinecite{garklaful00} can be processed
in the same way. The resulting expression for ${\cal H}$ up to the second order reads
%\marginpar{Hupto2}
% 
\begin{equation}\label{Hupto2}
{\cal H}^{(0+1+2)} \cong - \frac 12 \sum_{ij} \Phi_{ij}^{(2)} {\bf n}_i \cdot {\bf n}_j 
\end{equation}  
with
%\marginpar{Phidef}
% 
\begin{equation}\label{Phidef}
\Phi_{ij}^{(2)} = \tilde J_{ij} + \frac {\beta \tilde J_0} {2S} V_{ij}^{(1)} 
- \frac {(\beta \tilde J_0)^2} {6S^2} V_{ij}^{(2)}
\end{equation}  
and
%\marginpar{V2def}
% 
\begin{equation}\label{V2def}
V_{ij}^{(2)} \equiv \delta_{ij} \tilde J_0 - 3\tilde J_{ij} + \frac 3 {\tilde J_0 } 
\sum_l \tilde J_{il} \tilde J_{lj} - \frac 1 {\tilde J_0^2 } 
\sum_{ln} \tilde J_{il} \tilde J_{ln} \tilde J_{nj}.
\end{equation}  
In Fourier representation Eq.\ (\ref{Phidef}) takes the form
%\marginpar{PhiFourier}
% 
\begin{equation}\label{PhiFourier}
\Phi_{\bf k}^{(2)} = \tilde J_0 - (\tilde J_0 - \tilde J_{\bf k}) + 
\frac \beta {2S} (\tilde J_0 - \tilde J_{\bf k})^2 - 
\frac {\beta^2}{6S^2} (\tilde J_0 - \tilde J_{\bf k})^3.
\end{equation}  
It is not difficult to guess the form of ${\cal H}$ containing all orders of
$\beta \tilde J_0/S$ in the cumulant expansion at $T\ll \tilde J = JS^2$.
The result reads
%\marginpar{Hextrap}
% 
\begin{equation}\label{Hextrap}
{\cal H} \cong - \frac 12 \sum_{ij} \Phi_{ij} {\bf n}_i \cdot {\bf n}_j 
\end{equation}  
with 
%\marginpar{PhiExtrap}
% 
\begin{equation}\label{PhiExtrap}
\Phi_{\bf k} = \tilde J_0 - TS [1 - \exp(-\beta \varepsilon_{\bf k})],
\end{equation}  
where $\varepsilon_{\bf k} = (\tilde J_0 - \tilde J_{\bf k})/S = S(J_0 - J_{\bf k})$ is the
spin-wave spectrum of the ferromagnetic model.

One can generalize Eqs.\ (\ref{Hextrap}) and (\ref{PhiExtrap}) for the ferromagnetic model by adding the effect of a homogeneous field.
This is done by applying the same method to the field terms of the cumulant expansion of  Ref.\ \onlinecite{garklaful00}.
The result is
%
%\marginpar{HextrapH}
% 
\begin{equation}\label{HextrapH}
{\cal H} \cong {\cal H}_0 - {\bf h}^{\rm eff} \cdot \sum_i {\bf n}_i 
- \frac 12 \sum_{ij} \tilde J_{ij}^{\rm eff} {\bf n}_i \cdot {\bf n}_j,  
\end{equation}  
where
%
%\marginpar{H0def}
% 
\begin{equation}\label{H0def}
{\cal H}_0 = NTS\left( 1 - \frac 12 e^{-\beta H}\right)
- N \left(h + \frac {\tilde J_0} 2 \right),  
\end{equation}  
with the effective field and the exchange interaction given by
%
%\marginpar{heff}
% 
\begin{equation}\label{heff}
{\bf h}^{\rm eff} = TS \frac {\bf h} h ( 1 - e^{-\beta H}),
\qquad \tilde J_{\bf k}^{\rm eff} = TS e^{-\beta \varepsilon_{\bf k}} .
\end{equation}  
The spin-wave spectrum is here of the form
%
%\marginpar{SpectrumF}
% 
\begin{equation}\label{SpectrumF}
\varepsilon_{\bf k} = (h + \tilde J_0 - \tilde J_{\bf k})/S = H + S(J_0 - J_{\bf k}).
\end{equation}  
One can check that in the limit $S\to\infty$ the effective Hamilton function of 
Eq.\ (\ref{HextrapH}) reduces to the classical form of Eq.\ (\ref{ClassHam}).
The minimum of Eq.\ (\ref{HextrapH}) is attained for all spin vectors ${\bf n}_i$ directed
along ${\bf h}$ and is equal to 
%
%\marginpar{E0F}
% 
\begin{equation}\label{E0F}
E_0 = -Nh - \frac {N\tilde J_0} 2,
\end{equation}  
i.e., it coincides with the ground-state energy of the classical model.

It should be noted that although the form of ${\cal H}$ given by Eq.\ (\ref{HextrapH}) has been derived for $T\ll JS^2$, it yields qualitatively correct results for the thermodynamic functions in the whole range of temperatures.
For $T\gg JS^2$, deviations from the exact high-temperature asymptotes are or the relative order $1/S$, whereas the leading classical term is recovered (see Appendices \ref{app_spininfield} and \ref{app_twospins} for toy models).

Analogous calculations can be done for the antiferromagnetic model on bipartite lattices etc.
For the bipartite AFM model in zero field one obtains Eq.\ (\ref{Hupto2}) with
%\marginpar{PhiFourierAF}
% 
\begin{equation}\label{PhiFourierAF}
\Phi_{\bf k}^{(2)} = - \tilde J_{\bf k} + 
\frac \beta {2S} (\tilde J_0 + \tilde J_{\bf k})^2 - 
\frac {\beta^2}{6S^2} (\tilde J_0 + \tilde J_{\bf k})^2(\tilde J_0 + 2\tilde J_{\bf k}).
\end{equation}  
Unlike the case of Eq.\ (\ref{PhiFourier}), here it is difficult to guess a function which could yield an expansion of this form.
Thus another approach to this problem is needed.
The possibility of constructing the low-temperature form of the effective classical Hamilton function for the ferromagnetic model is an indication that an analytical method must exist which is not based on summing up the cumulant expansion and which works for both ferro- and antiferromagnets.

\section{Closed-form quasiclassical Hamiltonians with spin waves}
\label{sec_sw}

\subsection{A boson formalism}

Returning to the basic formula for the effective Hamilton function, Eq.\ (\ref{HClass}), we note that in many works the introduction of spin coherent states ${\bf n}_i$ for large-spin systems is followed by the assertment that such spins behave quasiclassically. 
In that case one can simply replace the spin operator ${\bf S}_i$ by $S{\bf n}_i$, which obviously leads to the classical Hamilton function.
Such an approach, however, misses quantum fluctuations of spins which are responsible for ${\bf S}_i \neq S{\bf n}_i$ and therefore possible corrections to the classical results.
To take quantum effects into account, we will proceed as follows.
First, we express spin operators in the
coordinate system defined by the vector ${\bf n}_i$ 
%\marginpar{SComp}
%
\begin{equation}\label{SComp}
{\bf S}_i = S_{iz} {\bf n}_i + S_{i+} {\bf n}_{i-} + S_{i-} {\bf n}_{i+}
\end{equation}
where
%
%\marginpar{nPerpDef}
\begin{equation}\label{nPerpDef}
{\bf n}_{i\pm} = ( {\bf n}_{ix} \pm i{\bf n}_{iy} )/2 .
\end{equation}
The vectors ${\bf n}_{ix}$ and ${\bf n}_{iy}$ are unit vectors which form, together with ${\bf n}_i \equiv {\bf n}_{iz}$, an orthogonal coordinate system.
Then we use the Holstein-Primakoff boson representation
%
%\marginpar{HolPrim}
\begin{eqnarray}\label{HolPrim}
&&
S_i^z = S - a_i^{+}a_i, \qquad S_i^{+} = (2S-a_i^+a_i)^{1/2}a_i,
\nonumber\\ 
&&
\qquad\qquad S_i^{-} = a_i^{+} (2S-a_i^+a_i)^{1/2}.
\end{eqnarray}
Here we expand square roots and keep in $\hat H$ only up to quadratic terms in boson operators.
A priori, this procedure is justified for $S\gg 1$.
In fact, however, it is similar to the usual approach made in the linear spin-wave theory, and in most cases it does not require large $S$. 
The latter thus becomes
%
%\marginpar{HamBoson}
\begin{equation}\label{HamBoson}
\hat H \cong {\cal H}^{(0)} + \hat H^{(1)} + \hat H^{(2)}.
\end{equation}
Here ${\cal H}^{(0)}$ is the classical Hamilton function given by Eq.\ (\ref{ClassHam}),
%
%\marginpar{HamBoson1}
\begin{equation}\label{HamBoson1}
\hat H^{(1)} = -\sqrt{\frac 2S} \sum_i [({\bf h}_i^{\rm MF} \cdot {\bf n}_{i+}) a_i^+ 
+ {\rm h.c.} ],
\end{equation}
where 
%
%\marginpar{hMFdef}
\begin{equation}\label{hMFdef}
{\bf h}_i^{\rm MF} \equiv {\bf h} +\sum_j \tilde J_{ij}{\bf n}_j
\end{equation}
is the classical molecular field, and 
%\marginpar{HamBoson2}
\begin{eqnarray}\label{HamBoson2}
&&
\hat H^{(2)} = -\frac 1S \sum_{ij} \tilde J_{ij} [({\bf n}_{i+} \cdot {\bf n}_{j-})
a_i^+ a_j + ({\bf n}_{i+} \cdot {\bf n}_{j+}) a_i^+ a_j^+ 
\nonumber\\
&&
\qquad {} + {\rm h.c.} ]  + \frac 1S \sum_i ({\bf h}_i^{\rm MF} \cdot {\bf n}_i) a_i^+ a_i.
\end{eqnarray}

The problem of finding the effective classical Hamilton function ${\cal H}$ for a spin  system, Eq.\ (\ref{HClass}), has been reduced, in this approximation, to calculating the average of an exponential of a quadratic form of boson operators over the boson vacuum.
The latter can be done analytically in general \cite{ber66} but the result for a many-spin system is so complicated that it has no practical significance.
On the other hand, in the most interesting temperature range $T\ll \tilde J = JS^2$ there is a strong short-range order, and thus the scalar products ${\bf n}_i \cdot {\bf n}_j$ are close to their zero-temperature values $\overline{ {\bf n}_i \cdot {\bf n}_j }$.
For this reason, we will replace in Eq.\ (\ref{HamBoson2}) ${\bf n}_{i+} \cdot {\bf n}_{j\pm}$ by $\overline{ {\bf n}_{i+} \cdot {\bf n}_{j\pm} }$, etc.
The advantage of it is that the $\overline{ {\bf n}_{i+} \cdot {\bf n}_{j\pm}}$  are translationally invariant and this allows us to rewrite the bosonic quadratic form in Fourier space as a sum over a {\em single} wave vector.
This greatly simplifies the solution.
We will also consider two toy models where the problem can be solved in a general, i.e., non-simplified form.
These are the spin-in-a-field model (Appendix \ref{app_spininfield}) and the two-spin model (Appendix \ref{app_twospins}).

The choice of the transverse vectors $ {\bf n}_{ix} $ and ${\bf n}_{iy}$ in Eq.\ (\ref{nPerpDef}) is not unique. 
One can rotate them in the plane perpendicular to ${\bf n}_i$, which does not
change the final results for physical quantities.
The quantities $\overline{ {\bf n}_{i+} \cdot {\bf n}_{j\pm} }$ in Eq.\ (\ref{HamBoson2}) can be specified already at the present stage by an appropriate choice of ${\bf n}_{ix}$ and ${\bf n}_{iy}$.
In most of magnetic models at zero temperature the vectors ${\bf n}_i$ lie in a plane. 
In that case it is convenient to choose $ {\bf n}_{ix} $ equal for all sites $i$ and perpendicular to this plane.
It follows automatically that ${\bf n}_{iy} = [ {\bf n}_i \times {\bf n}_{ix} ] $.
This yields
\begin{eqnarray}\label{nixnjx}
&&
\overline{ {\bf n}_{ix} \cdot {\bf n}_{jx} } = 1
\nonumber\\
&&
\overline{ {\bf n}_{iy} \cdot {\bf n}_{jy} } = \overline{ {\bf n}_i \cdot {\bf n}_j }
\end{eqnarray}
and thus
\begin{eqnarray}\label{npmnpm}
\overline{ {\bf n}_{i\pm}\cdot{\bf n}_{j\mp} } = 
( 1 + \overline{ {\bf n}_i\cdot{\bf n}_j } ) /4
\nonumber \\
\overline{ {\bf n}_{i\pm}\cdot{\bf n}_{j\pm} } = 
( 1 - \overline{ {\bf n}_i\cdot{\bf n}_j } ) /4.
\end{eqnarray}
Evidently the definitions above remain valid if the spin vectors in the ground state are collinear, as for ferromagnets or antiferromagnets in zero field or in strong fields.

Using the formulas obtained above in Eq.\ (\ref{HamBoson2}) and Fourier transforming one obtains
%
%\marginpar{HamBoson1Fou}
\begin{equation}\label{HamBoson1Fou}
\hat H^{(1)} = - \sum_{\bf k} 
( g_{\bf k} a_{\bf k} + g_{\bf k}^* a_{\bf k}^+ )
\end{equation}
with
%\marginpar{gdef}
\begin{equation}\label{gdef}
g_{\bf k} \equiv \sqrt{\frac 2 {SN}} \sum_i e^{i {\bf k} \cdot {\bf r}_i } 
({\bf h}_i^{\rm MF} \cdot {\bf n}_{i-})
\end{equation}
and
%
%\marginpar{HamBoson2Fou}
\begin{equation}\label{HamBoson2Fou}
\hat H^{(2)} = \sum_{\bf k} 
\left[ A_{\bf k} a_{\bf k}^+ a_{\bf k} + 
\frac 12 B_{\bf k} ( a_{\bf k}^+ a_{-\bf k}^+ + {\rm h.c.} )\right],
\end{equation}
where
%\marginpar{ABdef}
\begin{eqnarray}\label{ABdef}
&&
A_{\bf k} \equiv SJ_0' - S(J_{\bf k}'+J_{\bf k})/2 + \overline{{\bf H}\cdot {\bf n}_i},
\nonumber\\
&&
B_{\bf k} \equiv S(J_{\bf k}'-J_{\bf k})/2,
\end{eqnarray}
and
%\marginpar{Jkdef}
\begin{eqnarray}\label{Jkdef}
&&
J_{\bf k} = \sum_i e^{i {\bf k} \cdot ( {\bf r}_i - {\bf r}_j ) } J_{ij},
\nonumber\\
&&
J_{\bf k}' = \sum_i e^{i {\bf k} \cdot ( {\bf r}_i - {\bf r}_j ) } J_{ij}
\overline{ {\bf n}_i\cdot{\bf n}_j }.
\end{eqnarray}
In  Eq.\ (\ref{ABdef}) it has been assumed that in the ground state $\overline{{\bf H}\cdot {\bf n}_i}$ does not depend on $i$.
This is true for ferromagnets and bipartite antiferromagnets.

Now we are prepared to perform the main step of our method and to calculate the matrix
element over the boson vacuum:
%\marginpar{HClassBos}
%
\begin{equation}\label{HClassBos}
{\cal H} = {\cal H}^{(0)} - T\ln \langle 0_a | \exp [- \beta (\hat H^{(1)}+\hat H^{(2)})] | 0_a \rangle,
\end{equation}
cf.\ Eq.\ (\ref{HClass}).
Since bose operators with different ${\bf k}$ commute, 
$[a_{\bf k},a_{\bf q}^+] = \Delta({\bf k} - {\bf q})$, and $\hat H^{(2)}$ couples bosons with
${\bf k}$ and $-{\bf k}$, the exponential in Eq.\ (\ref{HClassBos}) can be factorized in a
product of exponentials containing different pairs of $a_{\bf k}$ and $a_{-\bf k}$.
Thus our many-body problem reduced to a problem of calculating the vacuum matrix element of
the exponential of a bilinear form in bose operators $a\equiv a_{\bf k}$ and 
$b\equiv a_{-\bf k}$ (see Appendix \ref{app_averaging}).
The result has the form
\begin{eqnarray}\label{AfterAvr}
&&
{\cal H} = {\cal H}^{(0)} + E_Q 
- \frac 14 \sum_{\bf k} (g_{\bf k}^*+g_{-\bf k})(g_{\bf k}+g_{-\bf k}^*)F_{\bf k}^+
\nonumber\\
&&\qquad\qquad
{} - \frac 14 \sum_{\bf k} (g_{\bf k}^*-g_{-\bf k})(g_{\bf k}-g_{-\bf k}^*) F_{\bf k}^-
\end{eqnarray}
where $\varepsilon_k=\sqrt{A_k^2-B_k^2}$ is the spin-wave spectrum,
\begin{equation}\label{Fpmdef}
F_{\bf k}^\pm \equiv \frac 1{A_{\bf k}\pm B_{\bf k}} \left[ 1-\frac{2T}{A_{\bf k}\pm B_{\bf k}+ \varepsilon_{\bf
k}\coth(\beta \varepsilon_{\bf k}/2)} \right]
\end{equation}
and the quantity
\begin{equation}\label{E0def}
E_Q = -\frac 12 \sum_{\bf k} \left[ A_{\bf k}-T\ln{\left(\cosh(\beta \varepsilon_{\bf k})+
\frac{A_{\bf k}}{\varepsilon_{\bf k}}\sinh(\beta \varepsilon_{\bf k})\right)} \right]
\end{equation}
will be shown to be related to the spin-wave quantum correction of the ground-state energy for antiferromagnets. 

Using Eqs.\ (\ref{gdef}) and (\ref{nPerpDef}) one can rewrite ${\cal H}$ as
\begin{equation}\label{AfterAvr1}
{\cal H} = {\cal H}^{(0)} + E_Q 
- \frac 1{2S} \sum_{ij} [ G_{ij}^{x} F_{ij}^+ + G_{ij}^{y} F_{ij}^- ], 
\end{equation}
where
\begin{eqnarray}\label{Gdef}
&&
G_{ij}^{x} \equiv ({\bf h}_i^{\rm MF} \cdot {\bf n}_{ix}) 
({\bf h}_j^{\rm MF} \cdot {\bf n}_{jx}) 
\nonumber\\
&&
G_{ij}^{y} \equiv ({\bf h}_i^{\rm MF} \cdot {\bf n}_{iy}) 
({\bf h}_j^{\rm MF} \cdot {\bf n}_{jy}) 
\end{eqnarray}
and
\begin{equation}\label{Fijdef}
F_{ij}^\pm \equiv \frac 1N \sum_{\bf k} e^{i{\bf k} ( {\bf r}_i - {\bf r}_j )} F_{\bf k}^\pm.
\end{equation}
The molecular field $ {\bf h}_i^{\rm MF} $ given by Eq.\ (\ref{hMFdef}) is a linear function of $\{{\bf n}_l\}$.
Therefore the quasiclassical Hamilton function ${\cal H}$ of Eq.\ (\ref{AfterAvr1}) contains {\em four-spin} interactions and is thus of a non-Heisenberg type.
Eq.\ (\ref{AfterAvr1}) can be simplified to a Heisenberg-type Hamilton function in the most interesting temperature range $T \ll \tilde J = JS^2$ by using the fact of a long- or strong short-range order at these temperatures.

Before proceeding to this, let us consider on passing the quantum corrections to the classical Hamilton function, ${\cal H}^{(0)}$, at high temperatures.
To the leading order in $1/T$ one obtains for $E_Q$
\begin{equation}\label{E0HighT}
E_Q \cong -\frac \beta 4 \sum_{\bf k} B_{\bf k}^2 = 
-\frac \beta {16S^2} \sum_{\bf k} (\tilde J_{\bf k}'-\tilde J_{\bf k})^2.
\end{equation}
Using $F_{\bf k}^\pm \cong \beta/2$ to leading order in $1/T$, as well as relations from the Appendix of Ref.\ \onlinecite{garklaful00}, one obtains for the remaining quantum corrections in Eq.\ (\ref{AfterAvr})
\begin{equation}\label{H1cum}
{\cal H}^{(1)} = - \frac \beta 2 \sum_{\bf k} g_{\bf k}^* g_{\bf k}
= - \frac \beta {4S} \sum_i [{\bf h}_i^{\rm MF} \times {\bf n}_i]^2.
\end{equation}
[The same result follows if one keeps only the linear boson terms in $\hat H$, i.e., by setting $A_{\bf k} = B_{\bf k} =0$ in Eq.\ (\ref{AfterAvr})].
It can be seen that ${\cal H}^{(1)}$ coincides with Eq.\ (20) of Ref.\
\onlinecite{garklaful00} without the last small term of order $\beta/S^2$.

To obtain a quasiclassical Hamilton function which is suitable in the low-temperature range, one has to make our approach self-consistent by applying a small-fluctuation approximation, as was already done above in order to simplify Eq.\ (\ref{HamBoson2}).
For three-dimensional systems with long-range order at low temperatures, this would be a straightforward task since one can write $ {\bf n}_i = {\bf n}_i^{(0)} + \delta {\bf n}_i $ and expand in small $\delta {\bf n}_i$ around the long-ranged ${\bf n}_i^{(0)}$.
For one- and two-dimensional systems without long-range order, there is no ${\bf n}_i^{(0)}$, yet deviations from the {\em short-range order} are small. 
Exploiting this smallness requires a more sophisticated and general approach which we are going to present now.

First, in Eq.\ (\ref{Gdef}) it is convenient to replace
\begin{equation}\label{Qdef}
{\bf h}_i^{\rm MF} \Rightarrow {\bf Q}_i \equiv 
{\bf h}_i^{\rm MF} - h_i^{\rm MF, 0} {\bf n}_i , 
\end{equation}
where $ h_i^{\rm MF, 0} $ is the zero-temperature value of $ h_i^{\rm MF} $. 
Since in the ground state spin vectors are pointing along the molecular field, ${\bf Q}_i$ deviates from zero by virtue of fluctuations only.
Thus in Eq.\ (\ref{Gdef}) which contains ${\bf Q}$ quadratically, one can neglect fluctuations of vectors $ {\bf n}_{ix} $ etc.
Second, we expand ${\bf Q}_i$ in the local basis of site $i$ as 
\begin{equation}\label{Qexp}
{\bf Q}_i = ( {\bf Q}_i \cdot {\bf n}_i ) {\bf n}_i 
+ ( {\bf Q}_i \cdot {\bf n}_{ix} ) {\bf n}_{ix}
+ ( {\bf Q}_i \cdot {\bf n}_{iy} ) {\bf n}_{iy}
\end{equation}
and perform the scalar product $ {\bf Q}_i \cdot {\bf Q}_j $, then we simplify it with the help of Eq.\ (\ref{nixnjx}).
Using the resulting formula, we eliminate $ ( {\bf Q}_i \cdot {\bf n}_{ix}) 
({\bf Q}_j \cdot {\bf n}_{jx}) $ in $G_{ij}^{x} $ in Eq.\ (\ref{Gdef}), after which Eq.\ (\ref{AfterAvr1}) becomes
\begin{equation}\label{QuasiHam}
{\cal H} = {\cal H}' + {\cal H}'' ,
\end{equation}
where
\begin{equation}\label{QuasiHamPr}
{\cal H}' = {\cal H}^{(0)} + E_Q 
- \frac 1{2S} \sum_{ij} F_{ij}^+  {\bf Q}_i \cdot {\bf Q}_j 
\end{equation}
and  
\begin{eqnarray}\label{QuasiHamDPr}
&&
{\cal H}'' = - \frac 1{2S} \sum_{ij} \big\{ 
\big( F_{ij}^- - \overline{ {\bf n}_i \cdot {\bf n}_j } F_{ij}^+ \big) 
( {\bf Q}_i \cdot {\bf n}_{iy} )( {\bf Q}_j \cdot {\bf n}_{jy} )
\nonumber\\
&& \qquad\qquad\quad
{} - F_{ij}^+ \big[ 
\overline{ {\bf n}_i \cdot {\bf n}_j } 
( {\bf Q}_i \cdot {\bf n}_i ) ( {\bf Q}_j \cdot {\bf n}_j )
\nonumber\\
&& \qquad\qquad\quad
{} + \overline{ {\bf n}_i \cdot {\bf n}_{jy} }
( {\bf Q}_i \cdot {\bf n}_i )( {\bf Q}_j \cdot {\bf n}_{jy} )
\nonumber\\
&& \qquad\qquad\quad
{} + \overline{ {\bf n}_{iy} \cdot {\bf n}_j }
( {\bf Q}_i \cdot {\bf n}_{iy} )( {\bf Q}_j \cdot {\bf n}_j )
\big] \big\}.
\end{eqnarray}
The advantage of representing ${\cal H}$ in the above form is that ${\cal H}''$ vanishes for models with collinear ground state.
Indeed, in this case $\overline{ {\bf n}_i \cdot {\bf n}_{jy} } = 0$.
Further, the quantity ${\bf Q}_i \cdot {\bf n}_i$ becomes {\em quadratic} in spin deviations and thus the term with $( {\bf Q}_i \cdot {\bf n}_i ) ( {\bf Q}_j \cdot {\bf n}_j )$ is of fourth order and can be neglected.
The remaining term in ${\cal H}''$ vanishes because of the factor $\big( F_{ij}^- - \overline{ {\bf n}_i \cdot {\bf n}_j } F_{ij}^+ \big)$ which can be shown to be zero for collinear states.
For ferromagnets, one has $\overline{ {\bf n}_i \cdot {\bf n}_j }=1$ and $F_{ij}^+ = F_{ij}^-$, while for antiferromagnets in zero field one has $\overline{ {\bf n}_i \cdot {\bf n}_j }=\pm 1$ and $F_{ij}^+ = \pm F_{ij}^-$ depending on whether $i$ and $j$ belong to the same or different sublattices, respectively.
Thus ${\cal H}''$ is nonzero only for the bipartite antiferromagnetic model in the field range $0 < H < H_{SF}$, where $H_{\rm SF}=2SJ_0$ is the spin-flip field, as well as for more complicated models which are not the subject of this paper.
Next we will consider the ferromagnetic and bipartite antiferromagnetic models separately.

\subsection{Quasiclassical Hamilton function for ferromagnets}

For the ground state of the ferromagnetic model one has $({\bf h} \cdot {\bf n}_i) = h$ and $({\bf n}_i \cdot {\bf n}_j) = 1$.
Thus the parameters defined in Eqs.\ (\ref{ABdef}) and (\ref{Jkdef}) simplify to
\begin{eqnarray}\label{ABsimpl}
&&
J_{\bf k}' = J_{\bf k}, \qquad B_{\bf k} = 0
\nonumber\\
&&
A_{\bf k} = S(J_0 - J_{\bf k}) + H = \varepsilon_{\bf k},
\end{eqnarray}
which results in $E_Q=0$ and
\begin{equation}\label{Fsimpl}
F_{\bf k}^+ = F_{\bf k}^- = F_{\bf k} = \frac T {\varepsilon_{\bf k}^2 }
(e^{-\beta \varepsilon_{\bf k}} + \beta \varepsilon_{\bf k} - 1).
\end{equation}
Thus Eq.\ (\ref{QuasiHam}) reduces to
\begin{equation}\label{HsimplF}
{\cal H} = {\cal H}' = {\cal H}^{(0)} - \frac 1{2S} \sum_{ij} F_{ij} {\bf Q}_i \cdot {\bf Q}_j.
\end{equation}
where
\begin{equation}\label{IdentsimplF}
{\bf Q}_i = {\bf h}_i^{\rm MF} - (h+\tilde J_0){\bf n}_i
\end{equation}
Recalling the definition of ${\bf h}_i^{\rm MF}$ given in Eq.\ (\ref{hMFdef}), one can write
\begin{eqnarray}\label{HtransfF}
&&
{\cal H} \cong {\cal H}^{(0)} - 
\frac 1 {2SN} \sum_{\bf k} F_{\bf k}\sum_{ijlm} e^{-i{\bf k} ( {\bf r}_i - {\bf r}_j )}
\nonumber\\
&&\qquad\qquad
{}\times [\delta_{il} ({\bf h} - h {\bf n}_i) + (\tilde J_{il} - \delta_{il}\tilde J_0) {\bf n}_l]
\nonumber\\
&&\qquad\qquad
{} \cdot
[\delta_{jm} ({\bf h} - h {\bf n}_j) + (\tilde J_{jm} - \delta_{jm}\tilde J_0) {\bf n}_m]
\nonumber\\
&&\quad
{} = {\cal H}^{(0)} - \frac S {2N} \sum_{\bf k} F_{\bf k}\sum_{ij} e^{-i{\bf k} ( {\bf r}_i - {\bf r}_j )}
\nonumber\\
&&\qquad\qquad
{} \times [{\bf H} - \varepsilon_{\bf k} {\bf n}_i ] \cdot 
[{\bf H} - \varepsilon_{\bf k} {\bf n}_j ]
\nonumber\\
&&\quad
{} = {\cal H}^{(0)} - \frac {NSH^2} 2 F_0 + HF_0\sum_i  ({\bf h}\cdot {\bf n}_i)
\nonumber\\
&&\qquad\qquad
{}- \frac S 2 \sum_{ij} [\varepsilon_{\bf k}^2 F_{\bf k}]_{ij} ({\bf n}_i \cdot {\bf n}_j),
\end{eqnarray}
where $[\varepsilon_{\bf k}^2 F_{\bf k}]_{ij}$ denotes Fourier transform of 
$\varepsilon_{\bf k}^2 F_{\bf k}$.
With the help of Eqs.\ (\ref{ClassHam}) and (\ref{Fsimpl}) this result can be brought into
the final form of Eq.\ (\ref{HextrapH}).

\subsection{Quasiclassical Hamilton function for antiferromagnets}

For the antiferromagnetic model it is convenient to replace $J \Rightarrow -J$ in all the formulas above.
The canting angle of a classical bipartite AF model in an applied field is given by
\begin{equation}\label{AngleAF}
\sin \alpha \equiv x = \frac H {2SJ_0} = \frac h {2\tilde J_0}
\end{equation}
for $h \leq 2\tilde J_0$ and $x=1$ for $h \geq 2\tilde J_0$.
Thus in the zero-temperature equilibrium state one has
\begin{eqnarray}\label{AvrAF}
&&
\overline{ {\bf H} \cdot {\bf n}_i } = xH
\nonumber\\
&&
\overline{ {\bf n}_i \cdot {\bf n}_j } = x^2 + (1-x^2)e^{-i{\bf b} ( {\bf r}_i - {\bf r}_j
)},
\end{eqnarray}
where ${\bf b}$ is the inverse lattice vector (for the simple cubic lattice ${\bf b} =
\{\pi, \pi, \pi\}$).
From Eq.\ (\ref{Jkdef}) one obtains $J_{\bf k}' = (2x^2-1)J_{\bf k}$, so that 
\begin{eqnarray}\label{ApmB}
&&
A_{\bf k} = S(J_0 + x^2 J_{\bf k})
\nonumber\\
&&
B_{\bf k} = S(1-x^2)J_{\bf k}
\nonumber\\
&&
A_{\bf k} + B_{\bf k} = S(J_0 + J_{\bf k})
\nonumber\\
&&
A_{\bf k} - B_{\bf k} = S(J_0 - J_{\bf k} + 2x^2 J_{\bf k}).
\end{eqnarray}
The antiferromagnetic spectrum has the form
\begin{equation}\label{SpectrAF}
\varepsilon_{\bf k} = 
S \sqrt{(J_0 + J_{\bf k})(J_0 - J_{\bf k} + 2x^2 J_{\bf k})}.
\end{equation}
In fact, this expression describes the two spin-wave branches of an antiferromagnet in field, i.e., the gapless $ \varepsilon_{1\bf k} = \varepsilon_{\bf b-k} = S \sqrt{(J_0 - J_{\bf k})(J_0 + J_{\bf k} - 2x^2 J_{\bf k})}$ and the gapped $\varepsilon_{2\bf k} = \varepsilon_{\bf k}$ one.
Whereas $\varepsilon_{2\bf k}$ describes spin motions within the plane specified by the equilibrium orientation of the two sublattices (this motion changes the Zeeman energy), the gapless $\varepsilon_{1\bf k}$ accounts for the motions out of plane, which cost no Zeeman energy.
The quantity ${\bf Q}_i$ defined by Eq.\ (\ref{Qdef}) is given by
\begin{equation}\label{IdentsimplAF}
{\bf Q}_i = {\bf h}_i^{\rm MF} - \tilde J_0{\bf n}_i
\end{equation}
[cf. Eq.\ (\ref{IdentsimplF})].
Now one can process ${\cal H}'$ of Eq.\ (\ref{QuasiHamPr}) in a way similar to that of ferromagnets in the preceding subsection, which yields 
\begin{eqnarray}\label{HtransfAF}
&&
{\cal H}' = {\cal H}^{(0)} + E_Q - \frac {NSH^2} 2 F_0^+ + 2SJ_0F_0^+\sum_i  ({\bf h}\cdot {\bf n}_i)
\nonumber\\
&&\qquad\qquad
{}- \frac S 2 \sum_{ij} [(A_{\bf k} + B_{\bf k})^2 F_{\bf k}^+]_{ij} ({\bf n}_i \cdot {\bf n}_j)
\end{eqnarray}
[cf. Eq.\ (\ref{HtransfF})].
With the help of Eqs.\ (\ref{ClassHam}) and (\ref{Fpmdef}) this equation can be brought into the final form
\begin{equation}\label{HeffAF}
{\cal H}' = {\cal H}_0 + E_Q - {\bf h}^{\rm eff} \cdot \sum_i {\bf n}_i 
+ \frac 12 \sum_{ij} \tilde J_{ij}^{\rm eff,+} {\bf n}_i \cdot {\bf n}_j,  
\end{equation}  
where $E_Q$ is given by Eq.\ (\ref{E0def}),
%
%\marginpar{H0AF}
% 
\begin{equation}\label{H0AF}
{\cal H}_0 = 
- N \left(\frac{ h (h-h^{\rm eff})}{4\tilde J_0} + \frac {\tilde J_0} 2 \right). 
\end{equation}  
The effective field and the exchange interaction have the form
%
%\marginpar{heffAF}
% 
\begin{equation}\label{heffAF}
{\bf h}^{\rm eff} = \frac { 2ST {\bf H} } { A_0 + B_0 + \varepsilon_0 
\coth (\beta \varepsilon_0/2) }
\end{equation}  
and
%
%\marginpar{JeffAF}
% 
\begin{equation}\label{JeffAF}
\tilde J_{\bf k}^{\rm eff,\pm} = \frac { 2ST (A_{\bf k} \pm B_{\bf k}) } 
{ A_{\bf k} \pm B_{\bf k} + \varepsilon_{\bf k} 
\coth (\beta \varepsilon_{\bf k}/2) }.
\end{equation}  
Expanding $\tilde J_{\bf k}^{\rm eff,+}$ for $\beta \varepsilon_{\bf k} \sim SJ/T \ll 1$ one recovers Eq.\ (\ref{PhiFourierAF}) in the zero-field case.
Clearly, it is very difficult to guess that the series in Eq.\ (\ref{PhiFourierAF}) can be
summed up to give  Eq.\ (\ref{JeffAF}).
It can be checked that in the classical limit Eq.\ (\ref{HeffAF}) goes over to the classical
Hamilton function ${\cal H}^{(0)}$.

We calculate next ${\cal H}''$ of Eq.\ (\ref{QuasiHamDPr}).
To this end, we define
\begin{equation}\label{xvecdef}
{\bf x} \equiv \frac {\bf h} {2\tilde J_0}
\end{equation}  
[cf. Eq.\ (\ref{AngleAF})] and use the explicit definition of the transverse vectors
\begin{eqnarray}\label{ntrans}
&&
{\bf n}_{ix} = \frac { [ {\bf n}_i \times {\bf x} ] } 
{ | [ {\bf n}_i \times {\bf x} ] | } e^{i{\bf b}{\bf r}_i}
= \frac { [ {\bf n}_i \times {\bf x} ] } 
{ x\sqrt{1-x^2} } e^{i{\bf b}{\bf r}_i}
\nonumber\\
&&
{\bf n}_{iy} = [ {\bf n}_i \times {\bf n}_{ix} ] = 
\frac { {\bf n}_i x^2 - {\bf x}  } 
{ x\sqrt{1-x^2} } e^{i{\bf b}{\bf r}_i}
\end{eqnarray}
(remember that we neglect fluctuations in ${\bf n}_i$ as all fluctuations are accounted for by ${\bf Q}_i$).
With these definitions and Eq.\ (\ref{AvrAF}) one has
\begin{eqnarray}\label{moreangles}
&&
\overline{ {\bf n}_i \cdot {\bf n}_{jy} } = -x\sqrt{1-x^2} 
\big[ 1 - e^{i{\bf b}( {\bf r}_i - {\bf r}_j } \big] e^{i{\bf b}{\bf r}_j}
\nonumber\\
&&
\overline{ {\bf x} \cdot {\bf n}_{iy} } = -x\sqrt{1-x^2} e^{i{\bf b}{\bf r}_i}.
\end{eqnarray}
Next we simplify the quantity $ {\bf Q}_i \cdot {\bf n}_i $ which can be rewritten in the following way 
\begin{eqnarray}\label{QnSimpl}
&&
{\bf Q}_i \cdot \tilde J_0{\bf n}_i = \frac 12 {\bf Q}_i \cdot ({\bf Q}_i
+ 2\tilde J_0{\bf n}_i) - \frac 12 {\bf Q}_i^2
\nonumber\\
&& \qquad\qquad 
{} \cong \frac 12({{\bf h}_i^{\rm MF}}^2 - \tilde J_0^2).
\end{eqnarray}  
Here we neglected the quadratic term ${\bf Q}_i^2$.
Substituting for ${\bf h}_i^{\rm MF}$ the expression given by Eq.\ (\ref{hMFdef}) one obtains
\begin{equation}\label{QnSimplRes}
{\bf Q}_i \cdot {\bf n}_i \cong -2\sum_l \tilde J_{il}( {\bf n}_l \cdot {\bf x} - x^2),
\end{equation}  
where ${\bf n}_{i}$ has vanished.
Now one can compute $ {\bf Q}_i \cdot {\bf n}_{iy} $ with the help of Eq.\ (\ref{ntrans}) and the formula 
\begin{equation}\label{Qx}
{\bf Q}_i \cdot {\bf x} = -\sum_l [\delta_{il}\tilde J_0 + \tilde J_{il}]( {\bf n}_l \cdot {\bf x} - x^2).
\end{equation}  
The result has the form
\begin{equation}\label{Qny}
{\bf Q}_i \cdot {\bf n}_{iy} \cong \frac { e^{i{\bf b}{\bf r}_i} } 
{ x\sqrt{1-x^2} }  \sum_l [\delta_{il}\tilde J_0 + (1-2x^2)\tilde J_{il}] ( {\bf n}_l \cdot {\bf x} - x^2) .
\end{equation}  
The expressions obtained above can be substituted into Eq.\ (\ref{QuasiHamDPr}), which after some algebra yields ${\cal H}''$ in the {\em anisotropic} Heisenberg form
\begin{equation}\label{QuasiHamDPrFin}
{\cal H}'' = - \frac 1 2 \sum_{ij} G_{ij} 
( {\bf n}_i \cdot {\bf x} - x^2)  ( {\bf n}_j \cdot {\bf x} - x^2),
\end{equation}  
where
\begin{equation}\label{GkRes}
G_{\bf k} = \frac { (1-x^2) J_{\bf k}^{\rm eff,+} 
+ x^2 J_{\bf b-k}^{\rm eff,+} - J_{\bf b-k}^{\rm eff,-} } 
{x^2 (1-x^2) } 
\end{equation}  
and $J_{\bf b-k}^{\rm eff,\pm}$ are defined by Eq.\ (\ref{JeffAF}).
It can be shown that $G_{\bf k}$ remains finite in both limits $x\to 0$ and $x\to 1$, and therefore, ${\cal H}''$ vanishes in these limits.

Let us consider the minimal value $E_0$ of the antiferromagnetic quasiclassical Hamilton function, Eq.\ (\ref{QuasiHam}).
Minimization of Eq.\ (\ref{QuasiHam}) yields the value of the canting angle which is the same as for the underlying classical model [cf.\ Eq.\ (\ref{AngleAF})]: 
\begin{equation}\label{hJratio}
x = \frac{ h^{\rm eff} }{  \tilde J_0^{\rm eff,+} } = \frac{ h }{ 2\tilde J_0 }.
\end{equation}  
For this $x$ the term ${\cal H}''$ vanishes. 
Using Eq.\ (\ref{AvrAF}) and noticing that $ \tilde J_{\bf b}^{\rm eff,+} = 0$ one obtains
\begin{equation}\label{E0AF}
E_0 = E_{\rm Cl} + E_Q, \qquad E_{\rm Cl} = -\frac {Nh^2}{4\tilde J_0} - \frac{N\tilde J_0}
2.
\end{equation}  
Here $E_{\rm Cl}$ is simply the ground-state energy of the classical model, as in the ferromagnetic case, cf. Eq.\ (\ref{E0F}).
The quantity $E_Q$ given by Eq.\ (\ref{E0def}) is of quantum origin and can be rewritten in the form
\begin{equation}\label{E0reexpr}
E_Q = \Delta E_0 - T  \sum_{\bf k} \ln ( n_{\bf k} + 1 ) 
+ \frac T 2 \sum_{\bf k} \ln \left[\frac {(ST)^2 } 
{ \tilde J_{\bf k}^{\rm eff,+} \tilde J_{\bf k}^{\rm eff,-} } \right],
\end{equation}  
where $n_{\bf k} \equiv 1/(e^{\beta\varepsilon_{\bf k}} -1)$ is the boson occupation number and
\begin{equation}\label{DeltaE0T0}
\Delta E_0 = -\frac 12 \sum_{\bf k} ( A_{\bf k} - \varepsilon_{\bf k} ).
\end{equation}  
is the first-order correction to the quantum ground-state energy.
In the limit $T\to 0$ one has $E_0 = E_{\rm Cl} + \Delta E_0 $.

Again, we stress that, although derived for $T\ll JS^2$, our effective quasiclassical Hamilton function ${\cal H}$ leads to qualitatively correct results in the whole temperature range, reproducing the leading classical terms in the high-temperature asymptotes of physical quantities.

\section{Illustration: spin-wave theory}
\label{sec_illustration}

For an illustration of the formalism developed above we consider a three-dimensional magnetic systems for which we know the thermodynamic functions derived by the standard spin-wave theory.
We want to demonstrate that classical SWT for the effective Hamilton function ${\cal H}$ yields quantum results.
Let us do it for the antiferromagnetic model.
Starting from the zero-temperature ordered state ${\bf n}_i^{(0)}$ one can write
\begin{eqnarray}\label{nLowT}
&&
{\bf n}_i = {\bf n}_i^{(0)} + \delta {\bf n}_i 
\nonumber\\
&&
\delta {\bf n}_i \cong {\bf n}_{ix}^{(0)} x_i + {\bf n}_{iy}^{(0)} y_i - 
{\bf n}_i^{(0)} (x_i^2 + y_i^2)/2,
\end{eqnarray}  
where $x_i, y_i \ll 1$.
Then ${\cal H}'$ of Eq.\ (\ref{HeffAF}) becomes a bilinear form of spin deviations $x_i, y_i$
\begin{eqnarray}\label{Hprbilin}
&&
{\cal H}' \cong E_0 + \frac  1 2 \sum_i h_i^{\rm MF, eff, 0} (x_i^2 + y_i^2)  
\nonumber\\
&&\quad
{} + \frac 12 \sum_{ij} \tilde J_{ij}^{\rm eff, +} ( x_i x_j + 
\overline{ {\bf n}_{iy} \cdot {\bf n}_{jy} } y_i y_j ) 
\end{eqnarray}  
where $ \overline{ {\bf n}_{iy} \cdot {\bf n}_{jy} } = \overline{ {\bf n}_i \cdot {\bf n}_j } $. 
Furthermore the energy $E_0$ is given by Eq.\ (\ref{E0AF}), and the effective molecular field is
\begin{equation}\label{HMFeff}
{\bf h}_i^{\rm MF, eff, 0} \equiv {\bf h}^{\rm eff} - \sum_j \tilde J_{ij}^{\rm eff, +} {\bf n}_j^{(0)}.
\end{equation}
For an antiferromagnet in an external field $h\leq 2\tilde J_0$ one has $h_i^{\rm MF, eff, 0} = 0$, as follows from Eq.\ (\ref{hJratio}) and $\tilde J_{\bf b}^{\rm eff, +} = 0$.
For ${\cal H}''$ of Eq.\ (\ref{QuasiHamDPrFin}) we obtain
\begin{equation}\label{Hdprbilin}
{\cal H}'' \cong -\frac 12 \sum_{ij} G_{ij} 
(\overline{ {\bf x} \cdot {\bf n}_{iy} })(\overline{ {\bf x} \cdot {\bf n}_{jy} }) y_i y_j .
\end{equation}
Adding Eqs.\ (\ref{Hprbilin}) and (\ref{Hdprbilin}) and using Eqs.\ (\ref{AvrAF}) and (\ref{moreangles}) yields ${\cal H}$ in the harmonic approximation
\begin{equation}\label{Hbilin}
{\cal H} \cong E_0 + \frac 12 \sum_{ij} [ \tilde J_{ij}^{\rm eff, +}  x_i x_j + 
 \tilde J_{ij}^{\rm eff, -}  y_i y_j ]
\end{equation}
After calculating the partition function and using Eqs.\ (\ref{E0AF}) and (\ref{E0reexpr}) one obtains the formula
\begin{equation}\label{lnZAF}
\ln {\cal Z} = {\rm const} - \beta (E_{\rm Cl} + \Delta E_0) 
+ \sum_{\bf k} \ln ( n_{\bf k} + 1 )   
\end{equation}
which is nothing else but the result of the spin-wave theory for quantum systems.

For the ferromagnetic model the considerations are much simpler.
One has ${\bf h}_i^{\rm MF, eff, 0} = TS$ and the harmonic form for ${\cal H}$ reads
\begin{equation}\label{HbilinF}
{\cal H} \cong E_0 + \frac {TS}2 \sum_{ij} [1-e^{-\beta\varepsilon_{\bf k}}]_{ij} 
( x_i x_j + y_i y_j ) .
\end{equation}
It yields an expression for $\ln {\cal Z}$ similar to that of Eq.\ (\ref{lnZAF}) with $E_0$ and $\varepsilon_{\bf k}$ appropriate for ferromagnets.

\section{Discussion}
\label{sec_discussion}

We have obtained an effective classical Hamilton function ${\cal H}$ for ferromagnets and bipartite antiferromagnets, which can be used to compute thermodynamic quantities of the original quantum spin system with the Hamiltonian $\hat H$.
Quantum effects are taken into account in ${\cal H}$ at the level of the linear spin-wave theory, which means that the results can be applied in all cases where the quantum corrections to the ground state are small and can be treated perturbatively.
Although the results for  ${\cal H}$ have been obtained for $T\ll JS^2$, the thermodynamic quantities computed with ${\cal H}$ remain qualitatively correct in the whole range of temperatures.
Indeed, since ${\cal H}$ recovers the classical Hamiltonian in the limit $S\to\infty$, the main classical terms in the high-temperature asymptotes are correctly reproduced and the errors are of relative order $1/S$.

Our results can be applied to one- and two-dimensional spin models which are characterized by the absence of long-range order.
Here a reduction of the quantum model to a classical one is an important step in simplifying the problem.
However, the main difficulty in low dimensions is due to long wavelength fluctuations which prevent ordering at any finite temperature, and this feature is characteristic to both quantum and classical systems. 
Thus solutions of effective classical models for quantum systems are not straightforward.
They require application of numerical methods such as Monte Carlo simulations or approximate analytical methods such as the $1/D$ expansion,\cite{gar94jsp,gar96jsp} where $D$ is the number of spin components.

The latter is very efficient for systems such as kagome and pyrochlore antiferromagnets which have a degenerate classical ground state and remain a classical spin liquids (i.e., they do not show any extended short-range order) down to zero temperature.
For these two models, even a $D\to\infty$ approximation yields excellent results for thermodynamic quantities, as compared to the numerical data.
Studying quantum effects in these systems with the help of effective classical Hamilton functions in combination with the $1/D$ expansion seems to be very promising.
However, for these models the form of ${\cal H}$ will be different and it is still to be determined.

In concluding, one should mention an alternative approach to the derivation of effective classical Hamilton functions which was developed by the Florence group (see Refs.\ \onlinecite{cucetal95,cuctogvervai00} and references therein), which, in combination with MC simulations, proved to be very efficient for two-dimensional antiferromagnets. 
The main difference between our and the Florence forms of ${\cal H}$ is that our approach leads to a long-range  effective exchange interaction $J_{ij}^{\rm eff}$ with a diverging range of interaction at $T\to 0$ [see, e.g.,  Eq.\ (\ref{heff})], whereas the expression for $J_{ij}^{\rm eff}$ of the Florence group remains restricted to nearest-neighbours for the quantum models with nn interactions while only the magnitude of $J$ is renormalized by quantum effects.
The latter makes it possible to solve one-dimensional ferro- and antiferromagnetic models by using the known solution for classical systems.
This leads to results that are in a good accord with the numerical data for the appropriate quantum models.\cite{cuctogvervai00}
In distinction, an analytical solution for the classical magnetic chain is not available for long-range interactions.

While the long-range form of $J_{ij}^{\rm eff}$ resulting from our method might be considered as a disadvantage from the practical point of view, we should stress that it appears in a natural way and has a physical explanation.
The integrals over the Brillouin zone in the usual SWT, such as the magnetization correction in ferromagnets $\Delta m = - \int d{\bf k} n_{\bf k}$, are at low temperatures restricted to the small-${\bf k}$ region satisfying $\beta \varepsilon_{\bf k} \lesssim 1$ because spin waves with larger ${\bf k}$ are frozen out.
In our effective classical formalism, the same cut-off is ensured by the effective interaction, i.e., by $\tilde J_{\bf k}^{\rm eff} = TSe^{-\beta \varepsilon_{\bf k}}$ for ferromagnets.
Clearly localization of $\tilde J_{\bf k}^{\rm eff}$ in the small-${\bf k}$ region at low temperatures means that the interaction is long ranged in the coordinate space.

\section*{Acknowledgment}

D.~G. thanks E. M. Chudnovsky for the warm hospitality and useful discussions during the last stage of this work at Lehman College--CUNY.  

\appendix

\section{One spin in a field}

\label{app_spininfield}

In this Appendix we shall apply our approach to the most simple model, i.e., one spin in a field:
\begin{equation}\label{appA1}
\hat H = - {\bf H} \cdot {\bf S}.
\end{equation}
In boson representation in the basis of spin coherent states the spin operator takes the form
\begin{equation}\label{appA2}
{\bf S} = (S-a^+a){\bf n} + \sqrt{2S}[{\bf n}_+a^+ +{\bf n}_-a]
\end{equation}
[cf. Eqs.\ (\ref{SComp}) and (\ref{HolPrim})].
The Hamiltonian of Eq.\ (\ref{appA1}) thus becomes:
\begin{equation}\label{appA3}
\hat H = - (S-a^+a) ({\bf H} \cdot {\bf n}) - \sqrt{2S}[({\bf H} \cdot
{\bf n}_+)a^+ +{\rm h.c.}].
\end{equation}

For this model we are able to calculate the effective classical Hamilton function
\begin{equation}\label{appA4}
{\cal H} = - T\ln \langle 0_a | \exp [-\beta\hat H] | 0_a \rangle
\end{equation}
at all temperatures, i.e., without assuming small fluctuations of the vector ${\bf n}$.
The corresponding matrix element over the boson vacuum is calculated easily resulting in:
\begin{eqnarray}\label{appA5}
&&\langle 0_a | \exp [- \beta (Aa^+a-g^*a^+-ga)] | 0_a \rangle
\nonumber\\ &&\qquad
= \exp \left[\frac{g^*g}{A^2} (e^{-\beta A}+\beta A-1)\right].
\end{eqnarray}
Therefore, the effective Hamilton function becomes:
\begin{equation}\label{appA6}
{\cal H} = - S({\bf H} \cdot {\bf n}) - \frac{ST} 2 \frac{[{\bf H} \times {\bf
n}]^2}
{({\bf H} \cdot {\bf n})^2}
\left[ e^{-\beta{\bf H} \cdot {\bf n}} + \beta {\bf H} \cdot {\bf
n}-1    \right].
\end{equation}
In the limit $T\gg H$ it reproduces the first three terms of the cumulant expansion  \cite{klafulgar99}
\begin{eqnarray}\label{appA7}
&&{\cal H} = - S({\bf H} \cdot {\bf n}) - \frac{\beta S}{4}[{\bf H}
\times {\bf n}]^2
\nonumber\\ &&\qquad
+ \frac{\beta^2 S}{12}({\bf H} \cdot {\bf n}) [{\bf H} \times {\bf 
n}]^2
+ O(\beta^3).
\end{eqnarray}
Unfortunately, at low temperatures Eq.\ (\ref{appA6}) leads to divergencies in the thermodynamic properties because the spin tends to direct itself against the field and to create a large number of bosons. 
This takes place since we have kept only quadratic boson terms for spin operators. 
To avoid such a divergence, we linearize ${\cal H}$ with respect to fluctuations near the zero-temperature state with the spin directed along the field. 
The linearized Hamiltonian has the form
\begin{equation}\label{appA8}
{\cal H} = -SH + TS(1-e^{-\beta H})\left(1-\frac{{\bf H} \cdot {\bf n}}{H}\right).
\end{equation}
Now one can easily calculate the partition function
\begin{equation}\label{appA9}
\ln{\cal Z} = {\rm const} + \beta SH +
\ln\left(\frac{1-e^{-2S(1-e^{-\beta H})}}
{1-e^{-\beta H}}\right),
\end{equation}
which is well behaved at low temperatures.
It correctly reproduces the exponential behaviour of the specific heat $C$. 
For large values of spin the deviation from the exact result is exponentially small:
\begin{eqnarray}\label{appA10}
&&C \cong \frac{H^2}{T^2} e^{-\beta H}\left(1-\frac{2S}{e^{2S}-1}\right) 
\nonumber\\
&&C_{\rm exact} \cong \frac{H^2}{T^2} e^{-\beta H}.
\end{eqnarray}
This approach gives also correctly the leading classical term of the specific heat at high temperature
\begin{equation}\label{appA11}
C \cong \frac{S(S+3)H^2}{3T^2},
\qquad C_{\rm exact} \cong \frac{S(S+1)H^2}{3T^2},
\end{equation}
but there is an error in the $1/S$ quantum correction.

%******************************************************************************

\section{Two spins}

\label{app_twospins}

Let us consider the model of two ferromagnetically coupled spins $S$:
\begin{equation}\label{appB1}
\hat H = - J {\bf S}_1 \cdot {\bf S}_2.
\end{equation}

First we rewrite the spin operators in terms of coherent states ${\bf
n},{\bf m}$
and two bosons $a_1^+,a_2^+$
\begin{eqnarray}\label{appB2}
&&
{\bf S}_1 \cong (S-a_1^+a_1){\bf n} + \sqrt{2S}[{\bf n}_+a_1^+ +{\bf
n}_-a_1]\nonumber\\
&&
{\bf S}_2 \cong (S-a_2^+a_2){\bf m} + \sqrt{2S}[{\bf m}_+a_2^+ +{\bf
m}_-a_2],
\end{eqnarray}
after which the Hamiltonian takes on the form:
\begin{eqnarray}\label{appB3}
&&
\hat H \cong {\cal H}^{(0)} + \hat H^{(1)} + \hat H^{(2)}
\nonumber\\
&&
\hat H^{(1)} =
- J\sqrt{2S^3}[({\bf n}_+ \cdot {\bf m}) a_1^+ + ({\bf n} \cdot {\bf
m}_+) a_2^+ + {\rm h.c.}]
\nonumber\\
&&
\hat H^{(2)} = JS({\bf n} \cdot {\bf m}) [a_1^+a_1 + a_2^+a_2]
\nonumber\\
&&
\qquad {} - JS [({\bf n}_+ \cdot {\bf m}_-) a_1^+a_2 + ({\bf n}_+ \cdot {\bf
m}_+) a_1^+a_2^+ + {\rm h.c.}].
\end{eqnarray}

After a canonical transformation to new boson operators
\begin{eqnarray}\label{appB5}
a_1^+ = (b_1^+ + b_2^+)/\sqrt{2},\qquad 
a_2^+ = (b_1^+ - b_2^+)/\sqrt{2}
\end{eqnarray}
the Hamiltonian becomes
\begin{eqnarray}\label{appB6}
&&
\hat H = {\cal H}^{(0)} + \hat H_1 + \hat H_2 \nonumber\\
&&
\hat H_1 = \frac B2 - \frac B2 (b_1^+ + b_1)^2 \nonumber\\
&&
\hat H_2 = -(g^*b_2^+ + gb_2)
+ Ab_2^+b_2 + \frac B2 (b_2^{+2} + b_2^2)
\end{eqnarray}
with
\begin{eqnarray}\label{appB7}
&&
A = \frac{JS} 2 (1+3{\bf n} \cdot {\bf m}), \qquad B = \frac{JS} 2 (1-{\bf n} \cdot {\bf m})
\nonumber\\
&&\qquad\qquad
g = iJS^{3/2}|[{\bf n}\times{\bf m}]| .
\end{eqnarray}

Since $\hat H_1$ and $\hat H_2$ commute one can perform independent vacuum averages for the two kinds of bosons.
For $\hat H_1$ one can use the formula:
\begin{equation}\label{appB8}
\langle 0_{b_1} | e^{x(b_1^+ + b_1)^2} | 0_{b_1} \rangle
= 1/\sqrt{1-2x},
\end{equation}
whereas the average with $\hat H_2$ can be calculated by a method similar to that presented in Appendix C.
This leads to the effective Hamilton function of the form:
\begin{eqnarray}\label{appB9}
{\cal H} = {\cal H}^{(0)} + \frac B2 + \frac T2 \ln(1-\beta B) + E_Q -
g^*g F^{-},
\end{eqnarray}
where $F^{-}$ is defined by Eq.\ (\ref{Fpmdef}) without the index ${\bf k}$.
The function $E_Q$  is given by Eq.\ (\ref{E0def})  without summation over  ${\bf k}$, and $\varepsilon = \sqrt{A^2-B^2}$.

Although this Hamilton function gives correctly the first two terms of a cumulant expansion, it breaks down at low temperatures.
Like a single spin in a field, at low temperatures the two spins  tend to orient antiferromagnetically with respect to each other and create a large number of bosons, which results in a divergence of the partition function.
In order to avoid this divergence, one has to expand ${\cal H}$ up to the quadratic terms in the relative deviation of the vectors ${\bf n}$ and ${\bf m}$, that is, one has to keep $1- {\bf n} \cdot {\bf m}$ and to neglect $(1- {\bf n} \cdot {\bf m})^2$ etc.
As a result the effective Hamilton function becomes:
\begin{equation}\label{appB10}
{\cal H} = -JS^2 + TS\frac{1-e^{-2\beta JS}}{2}(1- {\bf n} \cdot {\bf m}),
\end{equation}
which leads to the partition function
\begin{equation}\label{appB11}
\ln{\cal Z} = {\rm const} + \beta JS^2 +
\ln\left(\frac{1-e^{-S(1-e^{-2\beta JS})}}
{1-e^{-2\beta JS}}\right).
\end{equation}
At low temperatures one obtains an exponentially small specific heat, which differs from the exact result by an amount of order $1/S$:
\begin{eqnarray}\label{appB12}
&&
C \cong \frac{4J^2S^2}{T^2} e^{-2\beta JS} \left(1-\frac{S}{e^S-1}\right)
\nonumber\\
&&C_{\rm exact} \cong \frac{4J^2S^2}{T^2} e^{-2\beta JS}
\left(1-\frac{2}{4S+1}\right).
\end{eqnarray}
At high temperatures our approach gives correctly the leading 
classical term:
\begin{equation}\label{appB13}
C \cong \frac{J^2S^3(S+6)}{3T^2},
\qquad C_{\rm exact} \cong \frac{J^2S^2(S+1)^2}{3T^2},
\end{equation}
whereas there is an error in the $1/S$ correction.

%*****************************************************************************

\section{Vacuum average of the bilinear-boson-form exponential}

\label{app_averaging}

Here we calculate the exponential of a bilinear form in boson operators
over the boson vacuum:
\begin{equation} \label{appC1}
f(x)=\langle 0|e^{-x{\hat H}}|0 \rangle ,
\end{equation}
where the operator $\hat H$ has a form:
\begin{eqnarray} \label{appC2}
\hat H &=& A (a^+a + b^+b) + B (a^+b^+ + ba)
\nonumber\\
  &+& g_a^*a^+ + g_b^*b^+ + g_a a + g_b b .
\end{eqnarray}
First let us consider the following operators:
\begin{eqnarray} \label{appC3}
a^+(x)&=&e^{-x{\hat H}} a^+ e^{x{\hat H}} \nonumber\\
b(x) &=& e^{-x{\hat H}} b   e^{x{\hat H}}.
\end{eqnarray}
Differentiating with respect to $x$, we find that $a^+(x)$, $b(x)$ satisfy the system of equations:
\begin{eqnarray} \label{appC4}
&&\frac{da^+(x)}{dx} = -[{\hat H},a^+(x)] = -A a^+(x) -
Bb(x)- g_a
\nonumber\\
&&\frac{db(x)}{dx} = -[{\hat H},b(x)] = A b(x) + Ba^+(x)
+ g_b^* .
\end{eqnarray}
Integrating it, one obtains
\begin{eqnarray} \label{appC5}
a^+(x)  &=& ua^+ + vb + z_1   \nonumber\\
b(x) &=& u'b  + v'a^+ + z_1'^* ,
\end{eqnarray}
where $z_1=w_1g_a + w_2g_b^*$, $z_1'=w_1'g_a + w_2'g_b^*$,
\begin{eqnarray} \label{appC6}
\left( \begin{array}{cc}
u  & v \\
v' & u' \end{array} \right)
= \exp \left[ x \left(\begin{array}{cc}
-A & -B \\
B & A \end{array} \right)\right]
\end{eqnarray}
and
\begin{eqnarray} \label{appC7}
\left( \begin{array}{cc}
-w_1  & w_2  \\
-w'_2 & w'_1  \end{array} \right)
= \left( \begin{array}{cc}
u-1 & v \\
v'  & u'-1  \end{array} \right)
\left( \begin{array}{cc}
-A & -B \\
B & A \end{array} \right)^{-1} .
\end{eqnarray}
The explicit forms of $u,v,w_1$ and $w_2$ are
\begin{eqnarray}\label{appC8}
&&
u = \cosh(x\varepsilon) - A\frac{\sinh(x\varepsilon)}{\varepsilon}
\nonumber\\
&&
v = - B\frac{\sinh(x\varepsilon)}{\varepsilon}
\nonumber\\
&&
w_1 = A\frac{\cosh(x\varepsilon)-1}{\varepsilon^2}
- \frac{\sinh(x\varepsilon)}{\varepsilon} \nonumber\\
&&
w_2 = -B\frac{\cosh(x\varepsilon)-1}{\varepsilon^2}
\end{eqnarray}
with $\varepsilon =\sqrt{A^2-B^2}$.
The quantities $u',v',w'_1,w'_2$ are derived from $u,v,w_1,w_2$
by changing the sign of $x$ in Eqs.\  (\ref{appC8}).
For the pair of operators $b^+(x)$ and $a(x)$ one can obtain
the similar to Eqs.\ (\ref{appC5}) formulae:
\begin{eqnarray} \label{appC51}
b^+(x)  &=& ub^+ + va + z_2   \nonumber\\
a(x) &=& u'a  + v'b^+ + z_2'^*
\end{eqnarray}
with $z_2=w_1g_b + w_2g_a^*$, $z_2'=w_1'g_b + w_2'g_a^*$.

Before calculating $f(x)$ it is useful to introduce an auxiliary functions
$f_a(x)$ and $f_b(x)$
\begin{eqnarray}\label{appC9}
f_a(x) = \langle 0|e^{-x{\hat H}}a^+ |0 \rangle  \nonumber\\
f_b(x) = \langle 0|e^{-x{\hat H}}b^+ |0 \rangle.
\end{eqnarray}
After carrying the operators $a^+$ and $b^+$ past $e^{-x{\hat H}}$ in the
last terms using Eqs.\ (\ref{appC5},\ref{appC51}) one finds:
\begin{eqnarray}\label{appC10}
f_a(x) = \langle 0|a^+(x) e^{-x{\hat H}} |0 \rangle
       = v f_b^*(x) + z_1 f(x) \nonumber\\
f_b(x) = \langle 0|b^+(x) e^{-x{\hat H}} |0 \rangle
       = v f_a^*(x) + z_2 f(x)
\end{eqnarray}
and, therefore, one can express $f_a(x)$ and $f_b(x)$ through $f(x)$:
\begin{eqnarray}\label{appC11}
f_a(x) &=& \frac{z_1+vz_2^*}{1-v^2}f(x)\nonumber\\
f_b(x) &=& \frac{z_2+vz_1^*}{1-v^2}f(x) .
\end{eqnarray}

In order to find $f(x)$ we perform the following manipulations.
First, we take the derivative of $f(x)$:
\begin{eqnarray} \label{appC12}
-\frac{df}{dx} = \langle 0| e^{-x{\hat H}} {\hat H} |0\rangle
              &=& B\langle 0| e^{-x{\hat H}}a^+b^+ |0\rangle \nonumber\\
              &+&  g_a^*f_a(x)+g_b^*f_b(x)  .
\end{eqnarray}
Carrying $a^+b^+$ past $e^{-x{\hat H}}$ yields
\begin{eqnarray}\label{appC13}
&&-\frac{df}{dx} = g_a^*f_a(x)+g_b^*f_b(x) + B(uv+z_1z_2)f(x) + \nonumber\\
&&\quad
+ Bv[z_1f_a^*(x)+z_2f_b^*(x)] + B v^2 \langle 0|ab e^{-x{\hat H}}|0 \rangle .
\end{eqnarray}
Noticing from Eq.\ (\ref{appC12}) that
\begin{eqnarray}\label{appC14}
B \langle 0|a b e^{-x{\hat H}}|0 \rangle
= -\frac{df}{dx} - g_af_a^*(x)- g_bf_b^*(x) ,
\end{eqnarray}
one obtains the differential equation on $f(x)$
\begin{eqnarray}\label{appC15}
&&-(1-v^2)\frac{df}{dx} = g_a^*f_a(x)+g_b^*f_b(x) + B(uv+z_1z_2)f(x)\nonumber\\
&& + Bv[z_1f_a^*(x)+z_2f_b^*(x)] - v^2[g_af_a^*(x)+g_bf_b^*(x)] .
\end{eqnarray}
Substituting here Eq.\ (\ref{appC11}) for $f_a(x)$ and $f_b(x)$ 
and integrating over $x$, one finally arrives at
\begin{eqnarray}\label{appC16}
&&
T\ln{f(\beta)}=
A - T\ln{\left[\cosh{(\beta\varepsilon)}+\frac{A}{\varepsilon}\sinh{(\beta\varepsilon)}
\right]} \nonumber\\
&&
{} + \frac{(g_a^*+g_b)(g_a+g_b^*)}{2(A +B )}\left[ 1
-\frac{2T}{A +B +\varepsilon\coth(\beta\varepsilon/2)}\right]
\nonumber\\
&&
{} + \frac{(g_a^*-g_b)(g_a-g_b^*)}{2(A -B )}\left[ 1
-\frac{2T}{A -B +\varepsilon\coth(\beta\varepsilon/2)}\right] .
\end{eqnarray}

\vspace{-0.5cm}
%\bibliography{gar}

\end{document}